\documentclass[authoryear,preprint,review,12pt]{elsarticle}

\newcommand{\Rsun}{\mbox{$R_{\odot}$}}

\newcommand{\kms}{\mbox{km s$^{-1}$}}

\usepackage{graphics}

\journal{New Astronomy}

\begin{document}

\begin{frontmatter}



\title{NSVS\,07394765: A new low-mass eclipsing binary below 0.6\,M$_{\odot}$}


\author[OCakirli]{\"{O}m\"{u}r \c{C}ak{\i}rl{\i}\corref{cor1}}
\ead{omur.cakirli@gmail.com,Tel:+90 (232) 3111740, Fax:+90 (232) 3731403}
\cortext[cor1]{Corresponding author}

\address[OCakirli]{Ege University, Science Faculty, Astronomy and Space Sciences Dept., 35100 Bornova, \.{I}zmir, Turkey.}

\begin{abstract}
The multi-color photometric and spectroscopic\footnote{Based on observations obtained with 
the T{\"U}B{\.I}TAK National Observatory 1.5-meter telescope, which is owned and operated by the  
T{\"U}B{\.I}TAK.}  observations of the newly discovered eclipsing binary NSVS\,07394765 were obtained. The 
resultant light and radial velocities were analysed and the global parameters of the system:
T$_{1}$=3\,300\,K; T$_{2}$=3\,106\,K; M$_1$=0.36\,M$_{\odot}$; M$_2$=0.18\,M$_{\odot}$;
R$_1$=0.46\,R$_{\odot}$; R$_2$=0.50\,R$_{\odot}$; L$_{1}$=0.030\,L$_{\odot}$; L$_2$=0.026\,L$_{\odot}$; $i=89.2{^\circ}$; 
$a=5.97$\,R$_{\odot}$; $d=28$\,pc.
The chromospheric activity of its components is revealed by strong emission in the H$\alpha$ line and observed
flares. Empirical relations for mass-radius and mass-temperature are derived on the basis of the parameters of known binaries with 
low-mass dM components. 
\end{abstract}


\end{frontmatter}



\section{Introduction}\label{sec:intro}
The M dwarfs are the most numerous stars in our Galaxy. They are quite poorly investigated because of the selection effect. The values of masses, radii, luminosities and temperatures are less than those of 30 binaries with low-mass eclipsing binary components given by \citep{cakirli2012}. The mass-temperature and temperature-radii relation is determined by only a few low-mass stars. This situation has prevented the development of the models for the M dwarfs. It is created that all available models underestimate the radii (by around 10-15 per cent) and overestimate the temperatures (by 200-300 K) of short and long period binaries with M components \citep{cakirli2012}.

The Northern Sky Variability Survey (NSVS) contains a great number of photometric data \citep{wozniak04} 
that allows searching of variable stars and determination of their periods and types of variability. One of them 
was NSVS\,07394765 $\equiv$   2MASS J082551+242725 ($\alpha$=08$^{h}25^{m}51^{s}.3$, $\delta$=
+$24^{\circ}27^{\prime}05^{\prime\prime}.1$).

On the base of the NSVS photometry obtained in 1999--2000 we derived the ephemeris: 
HJD({{MinI}})=2451503.363 + 2.2656 $\times$ E and built its light curve (Fig.\,1).

\begin{figure}
 \centering
 \includegraphics[width=1.\columnwidth]{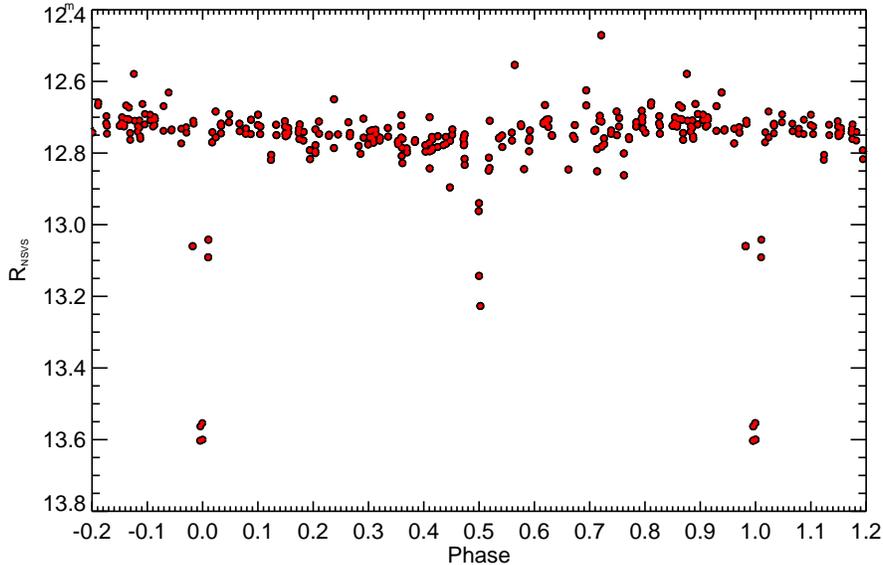}
\caption{NSVS photometry of NSVS\,07394765.}
\label{fig:nsvs}
\end{figure}

Initially NSVS\,07394765 attracted our interest by its very active components because there were 
only several systems with non-degenerate components and periods above the long-period limit 
of 4.26 days \citep{cakirli2012}: CU\,Cnc \citep{ribas03} with $P=2.794$ d,  
2MASS J01542930+0053266 \citep{becker08} with $P=2.619$ d, and T-Cyg1-12664 
\citep{cakirli2012} with $P=4.2631$ d. 

When we determined that the components of NSVS\,07394765 were low-mass eclipsing binary 
our interest increased and we undertook intensive photometric and spectral observations in 
order to determine its global parameters and to add a new information for the low-mass 
stars as well as for the very active and binary masses between 0.2-0.4\,M$_{\odot}$.

\begin{figure}
 \centering
\includegraphics[width=.8\columnwidth]{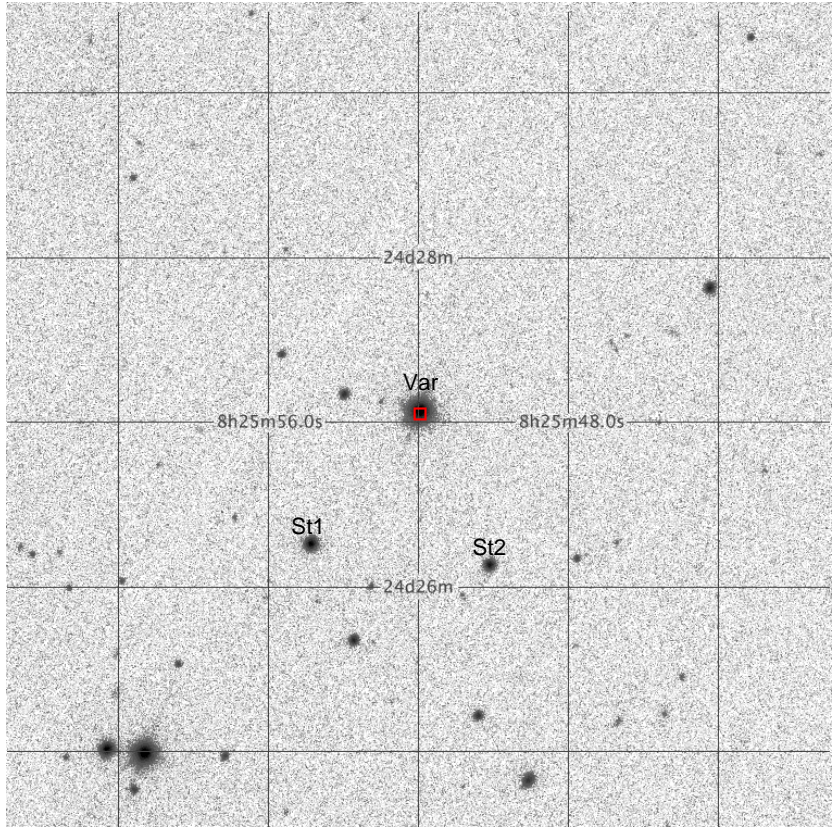}
 \caption{Observed field around NSVS\,07394765.}
\end{figure}

\section{Observations and data reduction}\label{sec:observation}
\subsection{Photometry}
NSVS\,07394765 was first identified in the {\sc Northern Sky Variability Survey} (NSVS; Wozniak 
et al. 2004) as a detached eclipsing binary system with a maximum, out-of-eclipse V-bandpass 
magnitude $R$=12$^m$.75 and a period of $P= 2.2656$ d. The data from the NSVS, obtained 
with the Robotic Optical Transient Search Experiment telescopes (ROTSE), contains positions, 
light curves and V magnitudes for about 14 million objects ranging in magnitudes from 8 to 15.5. 

The B, V, R, and I magnitudes for NSVS\,07394765 were  listed in the USNO NOMAD catalog 
as ( {\sc Naval Observatory Merged Astronomical Dataset}, NOMAD-1.0, \citep{zac04}, B=14$^m$.35, 
V=13$^m$.9, R=11$^m$.42 and I=11$^m$.22; on the other hand the infra-red magnitudes 
in three bandpasses were given as  J=10$^m$.554, H=9$^m$.880 and K=9$^m$.720 in the 
2MASS catalog \citep{cutri}.

The photometric observations of NSVS\,07394765 were carried out with the 0.4\,m telescope at the 
Ege University Observatory. The 0.4\,m telescope equipped with an Apogee CCD camera and standard Bessel 
BVRI bandpasses. The observations were performed on five nights between February\,01 and 
March\,30, 2009. To get the higher accuracy the target NSVS\,07394765 was placed near to the 
center of the CCD and three nearby stars located on the same frame were taken for comparison. The 
field of the variable and standard stars is shown in Fig.\,2. The stars TYC\,1941-1874-1 
and HD\,70897 were selected as comparison and check, respectively. Therefore the target and 
comparison stars could be observed simultaneously with an exposure time of 10 seconds. The 
differential observations of the comparison stars showed that they are stable during time span 
of our observations. The data were processed with standard data reduction procedures including 
bias and over scan subtraction, flat-fielding, and aperture photometry. The average uncertainty 
of each differential measurement was less than 0$^m$.030. The B-, V-, R- and I-bandpass magnitude 
differences, in the sense of variable minus comparison, are listed in Table\,1 (available in the 
electronic form at the CDS).

\begin{table*}
\scriptsize
\caption{Differential photometric measurements of NSVS\,07394765 in the B, V, R and I bandpasses.}
\setlength{\tabcolsep}{3.8pt} 
\begin{tabular}{cccccccc}
\hline
HJD(2\,400\,000+)  & $\Delta$B & HJD(2\,400\,000+) & $\Delta$V & HJD(2\,400\,000+) & $\Delta$R & HJD(2\,400\,000+) & $\Delta$I \\
\hline\hline
54520.53155			&1.214  &54520.52299	&0.659  & 54520.53481		& 0.259 & 54520.51694		& -0.179 \\
54520.53283			&1.220  &54520.52428	&0.647  & 54520.53610		& 0.282 & 54520.51823		&-0.185\\
...	&...	&... & ...&...	&...	&...	&...\\
...	&...	&... & ...&...	&...	&...	&...\\
\hline
\end{tabular}
\end{table*}

The light curve shows a deep primary eclipse with an amount of 0$^m$.85 in the V-bandpass 
and a shallow secondary eclipse with an amount of 0$^m$.45 which are clearly separated in 
phase, as is typical of fully detached binaries. The  primary and secondary eclipses occur 
almost 0.5 phase interval, indicating nearly circular orbit. An inspection of the nightly light 
curves presented in Fig.\,3 clearly indicates considerable out-of-eclipse light variations up 
to 0$^m$.2.  This intrinsic variation of the binary system manifests itself in the deeper primary eclipse. 

\begin{figure*}
 \centering
 \includegraphics[width=.8\columnwidth]{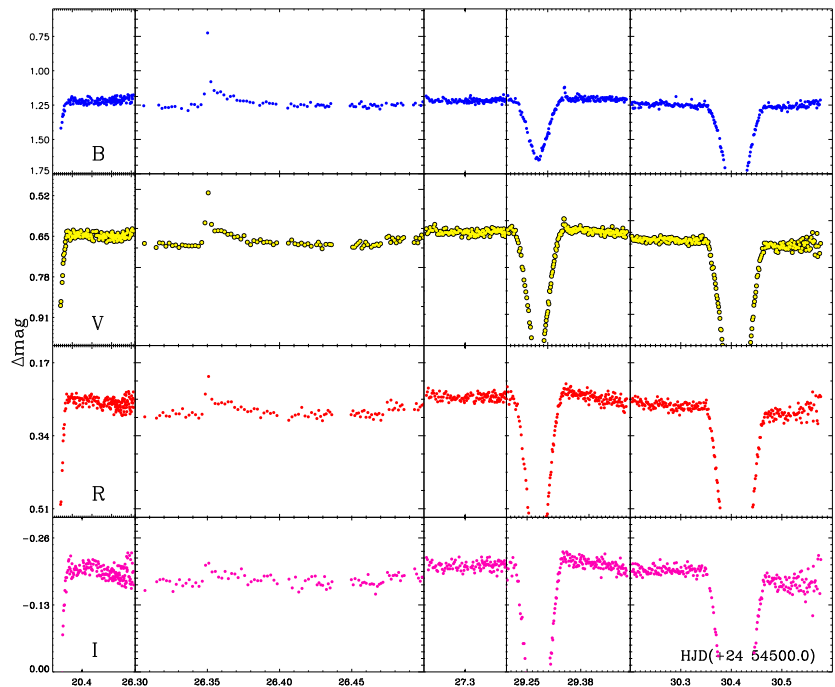}
\caption{The B-, V-, R-, and I-bandpass nightly light curves for NSVS\,07394765 from top to bottom. The
B-, V-, R-, and I-bandpass light curves clearly show that the brightness of the variable significantly 
varies from night to night, particularly in out of eclipse.}
\label{fig:nsvs}
\end{figure*}

\subsubsection{Orbital period and ephemeris}
The first orbital period for NSVS\,07394765 was determined as   $P=2.2656$\,d by \citet{wozniak04} 
from the NSVS database. Later on \citet{coug07} observed seven low-mass detached systems, 
including NSVS\,07394765. An orbital period and an initial epoch for the mid-primary eclipse were calculated using a 
least square fit. Partial primary and secondary eclipses which were detected in the time series 
photometric data were used in combination with the NSVS photometry to derive this 
ephemeris for the system. 

We obtained a time of mid-primary and secondary eclipse during our observing run. The 
mid-eclipse timings and their standard deviations are calculated using the method of 
\citet{kwee56}. These timings of the eclipses were listed in Table\,2 together with three 
primary and two secondary eclipses collected from literature. The times for mid-eclipses 
are the average of times obtained in four bandpasses. We  define the epoch of the system, 
T$_0$, to be the midpoint of the most complete primary eclipse. For this reason we use the 
B-, V-, R-, and I-bandpass data obtained on JD=2\,454\, 530 which cover almost the whole 
primary eclipse. A linear least square fit to the data listed in Table\,2 yields the new ephemeris as,

\begin{equation}
Min\, I\, (HJD)= 24\,54530.4143(5)+2.265754(11)\times E, 
\end{equation}  
where $E$ corresponds to the cycle number.  The residuals in the last column of Table\,2 are computed with the new ephemeris. While 
the orbital period is nearly the same with that determined by \citet{coug07} its uncertainty is now very smaller than 
estimated by them. In the computation of the orbital phase for individual observations we used this ephemeris.

\begin{table}
\caption{Times of minima of system.}
\begin{tabular}{crcc}
\hline\hline
HJD(2\,400\,000+)  & E	&Type          &O-C \\
\hline
51503.3625** 								  	&  -1336.0   	&     I       &       -0.0049 \\
51511.2973** 									&  -1332.5  	&     II      &       -0.0003 \\
51553.1954** 									&  -1314.0   	&     I       &       -0.0186 \\
51579.2599** 									&  -1302.5  	&     II      &       -0.0103 \\
53730.7303$^{\dagger}$ 				&   -353.0   	&     I       &        0.1271 \\
54529.2761$\pm$0.0005 				&    -0.5    	&     II      &       -0.0053 \\
54530.4094$\pm$0.0001 				&     0.0    	&     I       &       -0.0049 \\
\hline
\end{tabular}
\begin{list}{}{}
\item[$^{**}$]{\small From the NSVS database} 
\item[$^{\dagger}$]{\small \citet{coug07}} 
\end{list}
\end{table}

\subsection{Spectroscopy}
Optical spectroscopic observations of the NSVS\,07394765 were obtained with the Turkish 
Faint Object Spectrograph Camera (TFOSC)\footnote{http://tug.tug.tubitak.gov.tr/rtt150\_tfosc.php} 
attached to the 1.5 m telescope in January, 2012, under good seeing conditions. The wavelength 
coverage of each spectrum was 4000-9000 \AA~in 12 orders, with a resolving power of 
$\lambda$/$\Delta \lambda$ $\sim$7\,000 at 6563 \AA~and an average signal-to-noise ratio 
(S/N) was $\sim$120. We also obtained high S/N spectra of two M dwarfs  GJ\,740 
(M0\,V) and $\alpha$\,Cet (M1.5\,III) for use as templates in derivation of the radial velocities.

\section{Analysis}
\subsection{Effective temperature of the primary star}
We have used our spectra to reveal the spectral type of the primary component of NSVS\,07394765. For 
this purpose we have degraded the spectral resolution from 7\,000 to 3\,000, by convolving them with a 
Gaussian kernel of the appropriate width, and we have measured the equivalent widths of 
photospheric absorption lines for the spectral classification. We have followed the procedures of 
\citet{hernandez}, choosing helium lines in the blue-wavelength region, where the contribution 
of the secondary component to the observed spectrum is almost negligible. From several spectra 
we measured $EW_{\rm He I+ Fe I\lambda4922 }=0.98\pm 0.09$\,\AA,
$EW_{\rm Mg I\lambda5711}=0.54\pm 0.08$\,\AA~and  $EW_{\rm Ti II+ Fe II\lambda4203 }=1.38\pm 0.11$\,\AA.
From the calibration relations $EW$--Spectral-type of \citet{hernandez}, we have derived a spectral 
type of M2 with an uncertainty of about 1 spectral subclass. 

The effective temperature deduced from 
the calibrations of \citet{drill}, \citet{dejager}, \citet{Alonso}, \citet{flower} and \citet{popper} and 
\citet{Stra}  are 3\,300$\pm$90 K, 3\,488$\pm$150 K, 3\,382$\pm$180 K, 3\,060$\pm$150 K, 3\,100$\pm$300 K and 
3\,050$\pm$100 K, respectively. The standard deviations were estimated from the spectral-type 
uncertainty. The weighted mean of the effective temperature was obtained for the primary star 
as 3\,300$\pm$130 K.

Taking into account that $E(V-I)=$0.013, mag in the NSVS\,07394765 direction \citep{sch11}, we obtained its de-reddened
color index $(V-I)_0$=2.67 mag. According to the table 2 of \citet{one03}, this out-of-eclipse color index corresponds 
to a mean temperature of the binary T$_m$=3\,400\,K.

The catalogs USNO, NOMAD and GSC2.3 provide BVRIJHK 
magnitudes for NSVS\,07394765 with a few tenths of a magnitude uncertainties. Using the USNO 
B-mag of 14.35$\pm$0.20 and V-mag of 13.9$\pm$0.3 we obtained an observed color 
of B-V=0.45$\pm$0.36 mag. The observed infrared colors of J-H=0.674$\pm$0.021 and H-K=0.160$\pm$0.021 
are obtained using the JHK magnitudes given in the 2MASS catalog 
\citep{cutri}. These colors correspond to a main-sequence M1$\pm$2 star which is consistent 
with that estimated from the spectra.

\subsection{Radial velocity}
To derive the radial velocities of the components, the 12\,TFOSC spectra of the eclipsing binary 
were cross-correlated against the spectrum of GJ\,740, a single-lined M0\,V star, on an 
order-by-order basis using the {\sc fxcor} package in IRAF. The majority of the spectra 
showed two distinct cross-correlation peaks in the quadrature, one for each component 
of the binary. 

The heliocentric radial velocities for the primary (V$_p$) and the 
secondary (V$_s$) components are listed in Table\,3, along with the dates of observations 
and the corresponding orbital phases computed with the new ephemeris given in previous 
section. The radial velocities are plotted against the orbital phase in Fig.\,4.

\begin{table}
\caption{Heliocentric radial velocities of NSVS\,07394765. The columns give the heliocentric Julian date, the
orbital phase (according to the ephemeris in Eq.~1), the radial velocities of the two components with the 
corresponding standard deviations.}
\begin{tabular}{@{}ccccccccc@{}c}
\hline\hline
HJD 2400000+ & Phase & \multicolumn{2}{c}{Star 1 }& \multicolumn{2}{c}{Star 2 } 	\\
             &       & $V_p$                      & $\sigma$                    & $V_s$   	& $\sigma$	\\
\hline
55929.55435 &	0.5163	&  11.1	&  8.8	& ---	&  ---    \\
55929.64995 &	0.5585	&  19.9	&  8.7	& -25.5 &   9.8   \\
55930.42082 &	0.8987	&  25.6	&  4.4	& -51.2 &  11.1   \\
55930.46351 &	0.9176	&  17.8	& 11.1  & -41.1 &  12.2   \\
55930.55057 &	0.9560	&  14.3	& 10.9  & ---	&  ---    \\
55930.64022 &	0.9956	&  10.0	&  8.8	& ---	&  ---    \\
55931.25253 &	0.2658	& -22.2 &  3.5  &  99.9 &   6.7   \\
55931.39789 &	0.3300	& -18.2 &  4.3  &  88.4 &   5.4   \\
55931.44056 &	0.3488	& -17.3 &  4.3  &  81.1 &   9.8   \\
55931.48835 &	0.3699	& -13.3 &  4.4  &  77.7 &   7.8   \\
55932.37000 &	0.7590	&  37.4	&  5.2	& -88.7 &   6.5   \\
55932.49211 &	0.8129	&  33.3	&  3.2	& -82.1 &   7.4   \\
\hline \\
\end{tabular}
\end{table}

\begin{table}
\scriptsize
\caption{Results of the our ground-based BVRI light curves analyses for NSVS\,07394765. }
\setlength{\tabcolsep}{0.6pt} 
\begin{tabular}{lc}
\hline
Parameters  &$BVRI $ \\
\hline	
$i^{o}$											&89.23$\pm$0.03																								\\
T$_{eff_1}$ (K)								&3\,300[Fix]																											\\
T$_{eff_2}$ (K)								&3\,106$\pm$2    																								\\
$\Omega_1$									&13.595$\pm$0.051																							\\
$\Omega_2$									&7.110$\pm$0.019																								\\
r$_1$				   								& 0.0768$\pm$0.0003																							\\
r$_2$												& 0.0861$\pm$0.0003																							\\
$\frac{L_{1}}{(L_{1}+L_{2})}$ 	&0.582$\pm$0.003;0.562$\pm$0.002;0.541$\pm$0.002;0.522$\pm$0.003		\\
$\chi^2$											&0.9635																												\\		
Spot1														 								&Primary																		\\
										Latitude (deg)									&145																			\\
										Longitude (deg)								&290																			\\
										Angular radius (deg)						&20																				\\
										T$_{spot}$/T$_{photosphere}$		&0.89																			\\				             		     
Spot2																						&Primary																		\\
										Latitude (deg)									&90																				\\
										Longitude (deg)								&360																			\\
										Angular radius (deg)						&18																				\\
										T$_{spot}$/T$_{photosphere}$		&0.92																			\\
\hline
\end{tabular}
\end{table}

First we analysed the radial velocities for the initial orbital parameters. We used the 
orbital period held fixed and computed the eccentricity of the orbit, systemic velocity 
and semi-amplitudes of the radial velocities. The results of the analysis are as 
follows: $e$=0.001$\pm$0.001, i.e. formally consistent with a circular orbit, 
$\gamma$= 12$\pm$1 \kms, $K_1$=44$\pm$3 and $K_2$=88$\pm$4 \kms. Using 
these values we estimate the projected orbital semi-major axis and mass ratio 
as: $a$sin$i$=5.97$\pm$0.51 \Rsun~ and $q=\frac{M_2}{M_1}$=0.505$\pm$0.009

\begin{figure}
\includegraphics[width=10cm,angle=0]{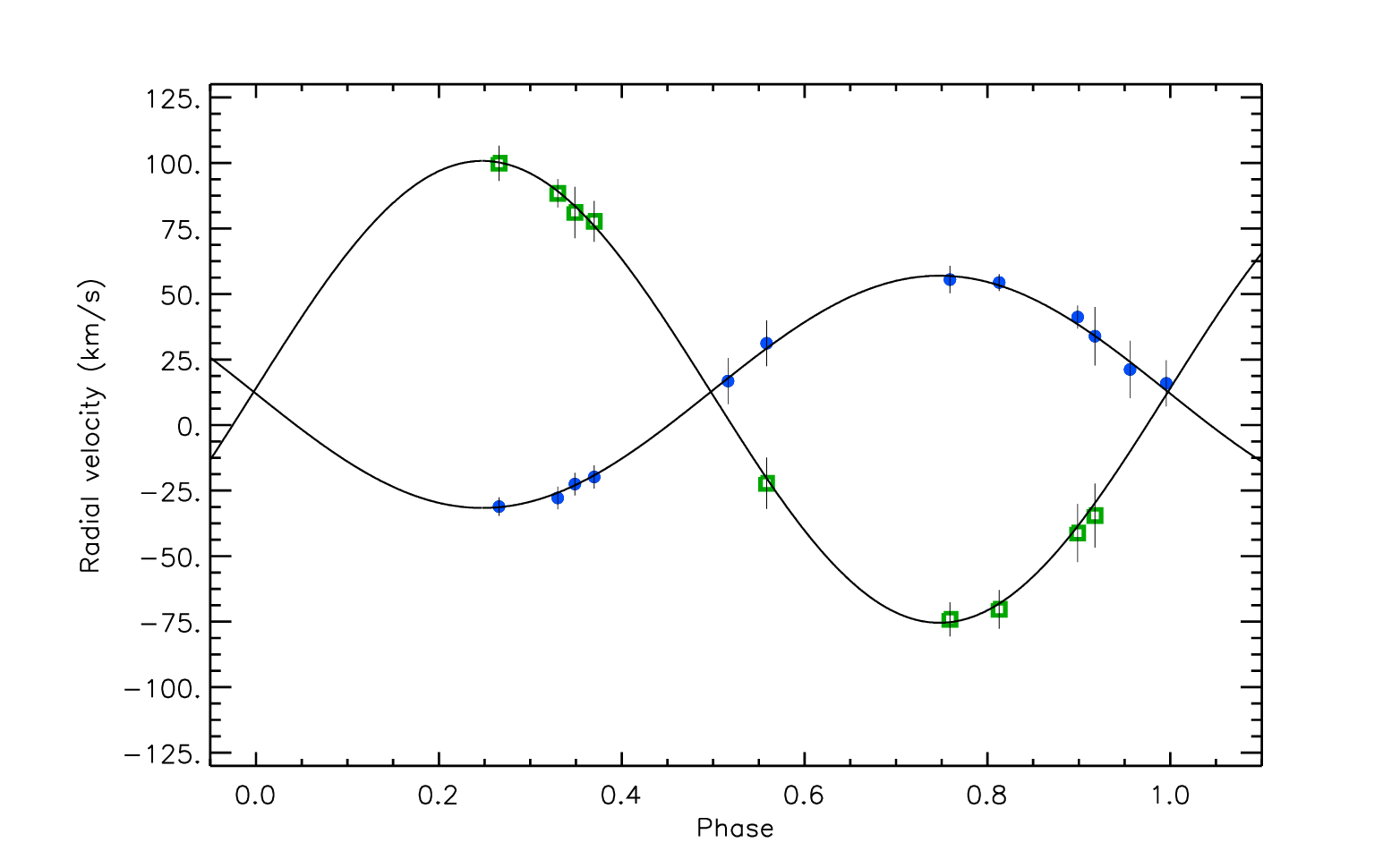}\\
\caption{Radial velocities folded on a period of 2.2656-day and the model. Points with error bars (error bars are masked by 
the symbol size in some cases) show the radial velocity measurements for the components of the system 
(primary: filled circles, secondary: open squares). } \end{figure}

\subsection{Light curve modeling}
We used the most recent version of the eclipsing binary light curve modeling algorithm of \citet{wil}, as 
implemented in the {\sc phoebe} code of Pr{\v s}a \& Zwitter (2005). The code needs some input 
parameters, which depend upon the physical properties of the component stars.

The BVRI photometric observations were analyzed simultaneously. The adjustable parameters in
the light curves fitting were the orbital inclination, the surface potentials, the effective temperature of 
secondary, and the luminosity of the hotter star. Our final results are listed in Table\,4 and the 
computed light curves (continuous line) are compared with the observations in Fig.\,5. The 
uncertainties assigned to the adjusted parameters are the internal errors provided directly 
by the Wilson-Devinney code. The  distortions at out-of-eclipses are clearly seen in the 
BVRI light curves. Therefore wave-like distortions are taken into account in the analysis of the BVRI light curves.

\section{Global parameters of the NSVS\,07394765}
The weighted mean of the orbital inclination and fractional radii of the components are found to 
be $i$=89.23$\pm$0.03, r$_1$=0.0768$\pm$0.0003, and r$_2$=0.0861$\pm$0.0003 from the light 
curves analyses. Using the orbital inclination and the semi-amplitudes of the radial velocities we 
determine the separation between the components as $a$=5.973$\pm$0.510 R$_{\odot}$. For the 
de-reddening, we used the $E$(B-V) = 0.009 value from \citet{sch11}. The interstellar reddening 
yield a distance to the system as 28$\pm$3 \,pc for the bolometric correction of -1.89 mag for a 
M2 main-sequence star \citep{drill}.  The effective temperature of 5\,770 K and bolometric magnitude 
of 4.74\,mag are adopted for the sun. The standard deviations of the parameters have been 
determined by JKTABSDIM\footnote{This can be obtained from http://http://www.astro.keele.
ac.uk/$\sim$jkt/codes.html} code, which calculates distance and other physical parameters 
using several different sources of bolometric corrections (Southworth et al. 2005). The best 
fitting parameters are listed in Table\,5 together with their formal standard deviations.

\begin{figure}
 \centering
 \includegraphics[width=.6\columnwidth,angle=90]{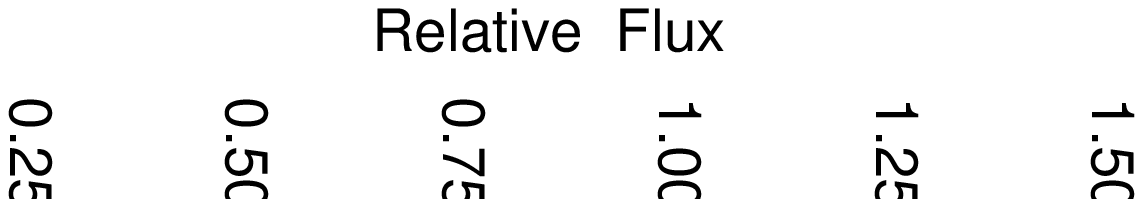}
 \caption{The observed and computed light curves of NSVS\,07394765.}
 \label{fig:chart}
\end{figure}

\begin{table}
\caption{Fundamental parameters of NSVS\,07394765.}
\begin{tabular}{lcc}
\hline
& \multicolumn{2}{c}{\hspace{0.25cm}NSVS\,07394765} 																									\\
Parameter 												& Primary	&	Secondary																						\\
\hline
Mass (M$_{\odot}$) 								& 0.360$\pm$0.005 & 0.180$\pm$0.004																\\
Radius (R$_{\odot}$) 								& 0.463$\pm$0.004 & 0.496$\pm$0.005																\\
$T_{eff}$ (K)											& 3\,300$\pm$200	& 3\,106$\pm$125      																\\
$\log~g$ ($cgs$) 									& 4.664$\pm$0.006 & 4.302$\pm$0.009																\\
$\log~(L/L_{\odot})$								& -1.53$\pm$0.06	& -1.59$\pm$0.08       															\\
$(vsin~i)_{calc.}$ (km s$^{-1}$)			&10$\pm$1  &11$\pm$1       																					\\   
Spectral Type											& M2V$\pm$1  	&M4V$\pm$1    																			\\
$a$ (R$_{\odot}$)									&\multicolumn{2}{c}{5.973$\pm$0.510}															\\
$V_{\gamma}$ (km s$^{-1}$)				&\multicolumn{2}{c}{12$\pm$1} 																		\\
$i$ ($^{\circ}$)										&\multicolumn{2}{c}{89.23$\pm$0.03	} 															\\
$q$															&\multicolumn{2}{c}{0.5049$\pm$0.0088}														\\
$d$ (pc)													& \multicolumn{2}{c}{28$\pm$3}																	\\
$\mu_\alpha cos\delta$, $\mu_\delta$(mas yr$^{-1}$) & \multicolumn{2}{c}{8$\pm$3, 48$\pm$5} 							\\
$U_o, V_o, W_o$ (km s$^{-1}$)  						& \multicolumn{2}{c}{-2$\pm$1, 15$\pm$1, 15$\pm$1}										\\ 
\hline  
 \end{tabular}
\end{table}

\subsection{Kinematics}
To study the kinematical properties of NSVS\,07394765, we used the system's centre-of-mass 
velocity, distance and proper motion values, which are given in Table\,5. The proper motion data
were taken from 2MASS catalogue \citep{cutri}, whereas the centre-of-mass velocity and distance 
are obtained in this study. The system's space velocity was calculated using \citet{sod87} algorithm. The 
U, V and W space velocity components and their errors were obtained and given in Table\,5. To obtain 
the space velocity precisely, the first-order galactic differential rotation correction was taken into account 
\citep{mih81}, and -1.08 and 0.65 \kms differential corrections were applied to U and V space velocity 
components, respectively. The W velocity is not affected in this first-order approximation. As for the 
LSR (Local Standard of Rest) correction, \citet{mih81} values (9, 12, 7)$_{\odot}$ \kms were used 
and the final space velocity of NSVS\,07394765 was obtained as  {\it S}\,=21 \kms. This value is in 
agreement with space velocities of the young stars.

\section{Activity of NSVS\,07394765}
The H$\alpha$ emission lines from the system is an useful indicator of chromospheric activity for late--type stars, in 
particularly M\,dwarfs. Typically, M\,dwarfs are divided into 4 subsets by \citet{stauffer86} in conformity with the strength 
of chromospheric activity. The lowest degree of chromospheric active M\,dwarfs have feeble H$\alpha$ absorption line in a 
spectrum. As the chromosphere increases the equivalent width of the H$\alpha$ absorption increases, then reducing and finally 
H$\alpha$ goes into the emission.

H$\alpha$ emission from the NSVS\,07394765 was present in every observed spectrum, as shown in Fig.\,6 as a 
function of the selected orbital phase. This is usual property of M\,dwarfs in general, because of the magnetic 
activity is frequently characterized by strong and variable H$\alpha$ line emission. Emission line characteristic 
varies with M\,dwarfs age, spectral type (particularly M2 or later), and the lifetime of magnetic activity \citep[]{westdr7,bell2012}. Like this variety also seen from the M\,dwarfs in close binary systems with other dwarf stars \citep[e.g.][]{dimitrov2010}.

We measured the H$\alpha$ equivalent width for each spectra and investigated variety of the strength of emissions.  Although it
seemed to change unusually in the range 3.2-4.4 \AA~ with orbital phase we noted a trend of the equivalent width to be higher level
of H$\alpha$ equivalent width at conjuncture.  

The H$\alpha$ structure was observed to be broader and stronger than other activity indicators line (e.g. Ca~{\sc i} absorption lines 
and H$\beta$ emission lines). Comparison of the structure with some binaries with low-mass M components from the \citet{cakirli2012} reveals the
strong H$\alpha$ emission of the system. This conclusion is not suprising taking into account the low temperature and fast rotation of its components. 
The mean value equivalent width 4 \AA~ of the H$\alpha$ emission of NSVS\,07394765 is considerably smaller than that of the accreting pre-main
sequence dMe stars which H$\alpha$ emission width has greater than ten angstrom.

Besides the variability of the line profiles there is information about optical flares of the system in the obtained photometrical data. Generally, flare activity is
typical for late-type stars. Until recently, the active M\,dwarfs in close binary system in the literature was 
CM\,Dra, V405\,And, YY\,Gem and GSC\,2314-0530 \citep[e.g.][]{dimitrov2010}. It should be noted that 2 observed flares occured 
around the HJD\,54526.35 and 54529.31. This involves correlation between the two signs of stellar activity: spots and flares. Both of them are appearances of the long--lived active area
on the components.

No other demonstrative clues for accretion, or outflow in the form of emission structure was observed. It does not appear that the H$\alpha$ emission is partially
coming from the third body or other physically processes, but more likely from many distributed active region on both stars. Longer duration and higher resolution spectroscopic 
monitoring would be necessary to accurately determine the geometry of the H$\alpha$ emission regions, and the active region timescales.

\begin{figure}
\centering
\includegraphics[width=0.6\columnwidth]{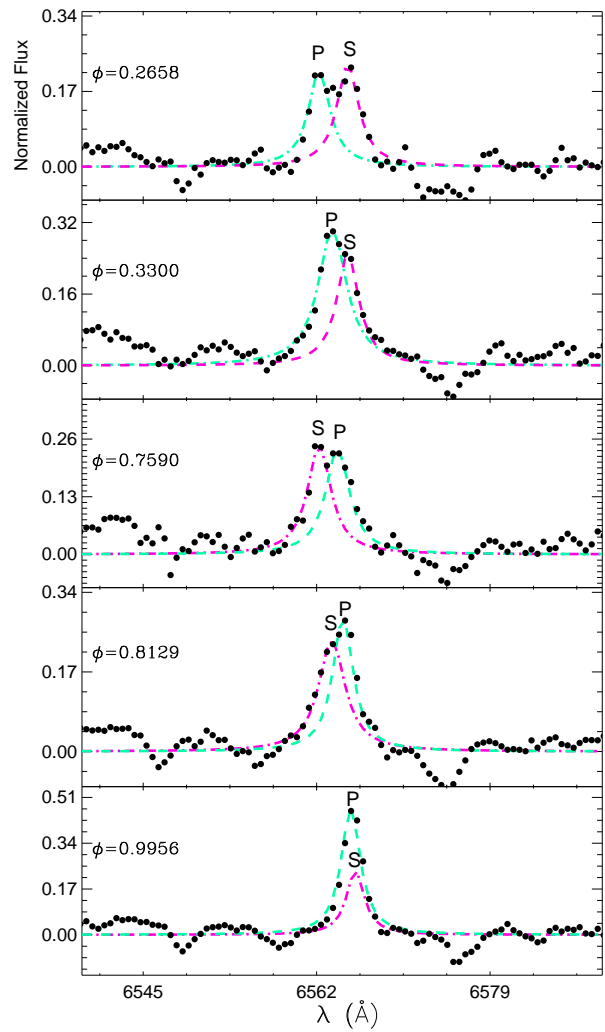}
\caption{Echelle spectra of H$\alpha$ lines are shown at several phases. Dashed lines shows the resulting fit of Gaussians to the line profile
with letters indicating each component. }
\label{fig:chart}
\end{figure}

\section{Conclusions}
The analysis of our photometric and spectral observations of the eclipsing binary NSVS\,07394765 allows us to derive
the following conclusions:

We have presented follow up photometric and first time spectroscopic observations of a low-mass binary, NSVS\,07394765, whose total mass of the components 
are below the limit of full convection dwarfs. The components of the system have masses typical M2 and M4, are in detached configuration. 

By simultaneous radial velocity solution and light curve solution we determined the global parameters of the system described in \S 4 and showed in Table\,5.

In Fig.\,7 we derived empirical positions of the NSVS\,07394765 components in the mass-radius (M-R) and mass - effective temperatures 
(M-T$_{\rm eff}$) planes relative to those of the well-determined low-mass dM components in the eclipsing binary systems. Theoretical 
M-R diagrams for a zero-age main-sequence stars with [M/H]=0 taken from the \citet{bar98} are also plotted to comparison.
Due to the high magnetic activity in the fast-rotating dwarfs their surfaces are covered by dark spot(s) or spot groups. Spot coverage in active dwarfs
yields larger radii and lower effective temperatures. 

The distorted light curve of NSVS\,07394765 were reproduced by two cool spots on the primary component. The next sign of the
activity of the system is the strong H$\alpha$ emission of its components. Moreover we registered 2 flares of NSVS\,07394765. Both of 
them occurred at the phases of maximum visibility of the larger stable cool spot on the primary.

The analysis of all appearances of magnetic activity revealed existence of long-lived active area on the primary of the system.
The high activity of the target is natural consequence of the fast rotation and low temperatures of its components.

Our study on one of the lowest-mass eclipsing binary NSVS\,07394765 presents a next small step toward understanding dMe
stars and adds a new information to the poor statistic of the low-mass dM stars. Recently they became especially interesting as
appropriate targets for planet searches due to the relative larger transit depths.

\begin{figure}
\centering
\includegraphics[width=.75\columnwidth]{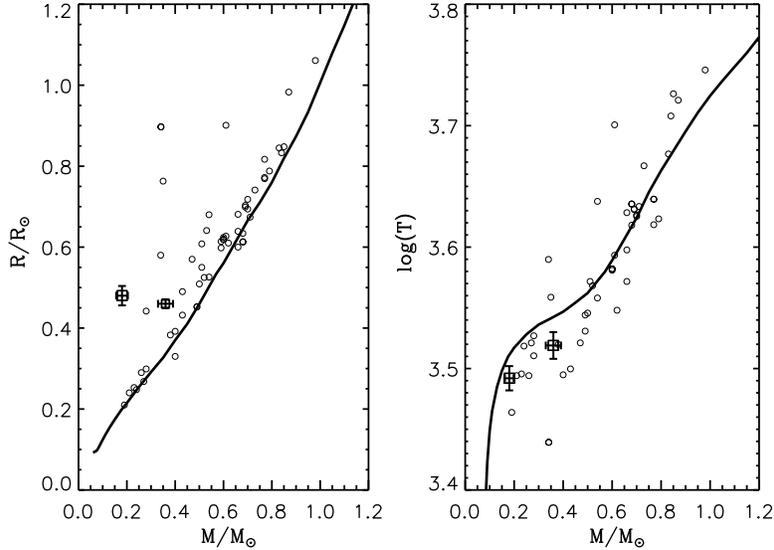}
\caption{Components of NSVS\,07394765 (squares with error bars) in the mass-radius plane (left panel). The less massive component is located
among the most deviated stars from the theoretical mass-radius relationship \citet{cakirli2012}. The lines show stellar evolution models from
Baraffe et al. (1998) for zero-age mian-sequence with [M/H]=0 (solid line). Components of the system (squares with error bars) in the mass-
effective temperature plane (right panel). }
\end{figure}

\section*{Acknowledgments}
We thank to T\"{U}B{\.I}TAK National Observatory (TUG) for a partial support in using RTT150 and T100 telescopes with project numbers 
10ARTT150-483-0, 11ARTT150-123-0 and 10CT100-101. 
We also thank to the staff of the Bak{\i}rl{\i}tepe observing station for their warm hospitality. 
%
%
The following internet-based resources were used in research for this paper: the NASA Astrophysics Data System; the SIMBAD database operated at CDS, Strasbourg, France; T\"{U}B\.{I}TAK 
ULAKB{\.I}M S\"{u}reli Yay{\i}nlar Katalo\v{g}u-TURKEY; and the ar$\chi$iv scientific paper preprint service operated by Cornell University.

\bibliographystyle{elsarticle-harv}

\begin{thebibliography}{00}
\bibitem[{{Alonso et al.} (1996)}]{Alonso}  Alonso, A., Arribas S., and Martinez-Roger, C., 1996, AA, 313, 873
\bibitem[\protect\citeauthoryear{Baraffe et al.}{1998}]{bar98} Baraffe I., Chabrier G., Allard F.,  Hauschildt P., 1998, A\&A, 337, 403 
\bibitem[\protect\citeauthoryear{Becker et al.}{2008}]{becker08}Becker A. et al., 2008, MNRAS, 386, 416
\bibitem[{{Bell} {et~al.}(2012){Bell}, {Hilton}, {Davenport}, {Hawley}, {West}, \& {Rogel}}]{bell2012} {Bell}, K.~J., {Hilton}, E.~J., {Davenport}, J.~R.~A., {Hawley}, S.~L., {West},  A.~A., \& {Rogel}, A.~B. 2012, PASP
\bibitem[\protect\citeauthoryear{Cakirli, \.{I}banoglu \& Sipahi}{2012}]{cakirli2012}Cakirli O., {\.I}banoglu C., \& Sipahi S., 2012, MNRAS, in press
\bibitem[\protect\citeauthoryear{Coughlin \& Shaw}{2007}]{coug07}Coughlin J., Shaw J., 2007, J. of Southeastern Assoc. for Res. in Astr., 1, 7
\bibitem[{{Cutri et al.} (2003)}]{cutri} Cutri R. M., et al., 2003, The IRSA {\em 2MASS} All-Sky Point Source Catalog, NASA/IPAC Infrared Science Archive.~http://irsa.ipac.caltech.edu/applications/Gator/ 
\bibitem[{{Drilling \& Landolt} (2000)}]{drill} Drilling J. S., Landolt A. U., 2000, Allen's astrophysical quantities, 4th ed. Edited by Arthur N. Cox. ISBN: 0-387-98746-0. Publisher: New York: AIP Press; Springer, 2000, p.381
\bibitem[{{Dimitrov} \& {Kjurkchieva}(2010)}]{dimitrov2010}{Dimitrov}, D.~P., \& {Kjurkchieva}, D.~P. 2010, MNRAS, 406, 2559
\bibitem[{{Flower} (1996)}]{flower} Flower P. J., 1996, ApJ, 469, 355
\bibitem[{{Hern\'andez et al.} (2004)}]{hernandez}  Hern\'andez J., Calvet N., Brice\~no C., Hartmann L., Berlind P., 2004, AJ, 127, 1682
\bibitem[{{Johnson \& Soderbloms}(1987)}]{sod87} Johnson, D. R. H., \& Soderbloms, D. R., 1987, AJ, 93, 864 
\bibitem[{{de Jager \& Nieuwenhuijzen} (1987)}]{dejager} de Jager C., Nieuwenhuijzen H., 1987, AA, 177, 217
\bibitem[Kwee \& van Woerden (1956)]{kwee56} Kwee K. K. \& van Woerden H., 1956, BAN, 12, 327
\bibitem[{{Mihalas \& Binney}(1981)}]{mih81} Mihalas, D., \& Binney, J., 1981. in Galactic Astronomy, 2nd edition, Freeman, San Fransisco, p.181
\bibitem[\protect\citeauthoryear{Mullan \& MacDonald}{2001}]{mullan01} Mullan D.~J., MacDonald J., 2001, ApJ, 559, 353 
\bibitem[\protect\citeauthoryear{Pr\'sa \& Zwitter}{2005}]{prsa05}Pr\'sa A., Zwitter T., 2005, ApJ, 628, 426
\bibitem[{{Popper} (1980)}]{popper} Popper D. M., 1980, Ann. Rev. AA, 18, 115
\bibitem[\protect\citeauthoryear{Straizys \& Kuriliene}{1981}]{Stra} Straizys V., Kuriliene G., 1981, Ap\&SS, 80, 353 
\bibitem[Southworth et al.(2005)]{sout05} Southworth J., Smalley B., Maxted P. F. L., Claret A. \& Etzel P. B. 2005, MNRAS, 363, 529
\bibitem[\protect\citeauthoryear{Ribas}{2003}]{ribas03}Ribas I., 2003, A\&A, 398, 239
\bibitem[\protect\citeauthoryear{Schlafly \& Finkbeiner}{2011}]{sch11} Schlafly E.~F., Finkbeiner D.~P., 2011, ApJ, 737, 103 
\bibitem[\protect\citeauthoryear{Stauffer \& Hartmann}{1986}]{stauffer86}Stauffer J.R., Hartmann L.W., 1986, ApJS, 61, 531
\bibitem[{{West} {et~al.}(2011){West}, {Morgan}, {Bochanski}, {Andersen},  {Bell}, {Kowalski}, {Davenport}, {Hawley}, {Schmidt}, {Bernat}, {Hilton}, {Muirhead}, {Covey}, {Rojas-Ayala}, {Schlawin}, {Gooding}, {Schluns},
  {Dhital}, {Pineda}, \& {Jones}}]{westdr7} {West}, A.~A., {et~al.} 2011, AJ, 141, 97
\bibitem[Wilson \& Devinney (1971)]{wil} Wilson R.E. \& Devinney E.J., 1971, ApJ, 166, 605
\bibitem[\protect\citeauthoryear{Wozniak et al.}{2004}]{wozniak04}Wozniak P.R., Vestrand C.W., Akerlof R., et al., 2004, AJ, 127, 2436
\bibitem[\protect\citeauthoryear{VandenBerg \& Clem}{2003}]{one03} VandenBerg D.~A., Clem J.~L., 2003, AJ, 126, 778 
\bibitem[\protect\citeauthoryear{Zacharias, Monet \& Levine}{2004}]{zac04}Zacharias N., Monet D. G., Levine S. E., 2004, AAS, 205, 4815Z



\end{thebibliography}



\end{document}